\documentclass[journal=jacsat,manuscript=article,onecolumn]{achemso}

\usepackage{chemformula} 
\usepackage[T1]{fontenc} 
\usepackage{amsmath}
\SectionNumbersOff

\author{Pablo Tieben}
\affiliation{Leibniz University Hanover, Institute of Solid State Physics, Appelstrasse 2, 30167 Hannover, Germany}
\alsoaffiliation[Second University]{Phyikalisch-Technische Bundesanstalt, Bundesallee 100, 38116 Braunschweig, Germany}
\email{tieben@qute.uni-hannover.de}
\author{Jan Rhensius}
\affiliation{QZabre AG, 8050 Zurich, Switzerland}
\author{Takuya F. Segawa}
\affiliation{Institute of Molecular Physical Science / Laboratory of Physical Chemistry,Department of Chemistry and Applied Biosciences, ETH Zurich, 8093 Zurich, Switzerland}
\author{Risei Abe}
\affiliation{Chitose Institute of Science and Technology, 758-65 Bibi, Chitose, Hokkaido 066-8655, Japan}
\author{Konosuke Shimazaki}
\affiliation[Kyoto]{Department of Electronic Science and Engineering, Kyoto University, Kyoto 615-8510, Japan}
\alsoaffiliation[Kyoto2]{Photonic Quantum Sensing Science and Engineering Center, Graduate School of Engineering, Kyoto University, Kyoto 615-8510, Japan}
\author{Shigeki Takeuchi}
\affiliation[Kyoto]{Department of Electronic Science and Engineering, Kyoto University, Kyoto 615-8510, Japan}
\alsoaffiliation[Kyoto2]{Photonic Quantum Sensing Science and Engineering Center, Graduate School of Engineering, Kyoto University, Kyoto 615-8510, Japan}
\author{Andeas W. Schell}
\affiliation{Institute of Semiconductor and Solid State Physics, Johannes Kepler University, Altenberger Strasse 69, Linz 4040, Austria}
\alsoaffiliation[Second University]{Leibniz University Hanover, Institute of Solid State Physics, Appelstrasse 2, 30167 Hannover, Germany}
\alsoaffiliation[Third University]{Phyikalisch-Technische Bundesanstalt, Bundesallee 100, 38116 Braunschweig, Germany}
\author{Hideaki Takashima}
\affiliation[CIST]{Chitose Institute of Science and Technology, 758-65 Bibi, Chitose, Hokkaido 066-8655, Japan}
\alsoaffiliation[Kyoto]{Department of Electronic Science and Engineering, Kyoto University, Kyoto 615-8510, Japan}
\alsoaffiliation[Kyoto2]{Photonic Quantum Sensing Science and Engineering Center, Graduate School of Engineering, Kyoto University, Kyoto 615-8510, Japan}

\title{Bright Single-Photon Emission from Individual Tin-Vacancy Centers in Multi-Cone Diamond Waveguides}

\begin{document}

\begin{abstract}
Diamonds containing color centers have recently gathered significant attention for photonic quantum technologies, including quantum sensing, photonic quantum computers, and quantum networks. Among the various color centers, tin\nobreakdash-vacancy (\ch{SnV}) centers are particularly promising due to the high emission efficiency from the zero\nobreakdash-phonon line and due to their long spin coherence times. However, the extraction of photons from diamond remains a key challenge. Here we demonstrate high photon extraction from a single \ch{SnV} center incorporated in a diamond nanopillar with tapered sidewalls and a multi-cone structure. A sharp emission peak with a full width at half maximum (FWHM) of $6$\,nm was observed at a wavelength of $619$\,nm. Furthermore, the second\nobreakdash-order correlation function exhibited an antibunching dip well below $g^{(2)}(0) = 0.5$, indicating single\nobreakdash-photon emission. Remarkably, the emitter achieved a high saturation count rate of approximately $9$\,Mcps. These results establish our nanopillar platform as a promising candidate for bright and stable quantum sources and sensors based on \ch{SnV^{-}} centers in diamond.
\end{abstract}

%
%
\section{Introduction}
Color centers in diamond have attracted attention in recent years for photonic quantum technologies, including quantum sensing \cite{Maze2008,Maletinsky2012a,Kucsko2013, Alkahtani2018,Segawa2023}, photonic quantum computers \cite{Wrachtrup2006,Xiang2013a}, and quantum networks \cite{Kimble2008, Bersin2024, Ruf2021, Borregaard2020}. Among these, nitrogen\nobreakdash-vacancy (\ch{NV}) centers have been the most studied across various areas \cite{Jelezko2006, Schroder2016}. However, a key limitation of \ch{NV} centers is their low zero\nobreakdash-phonon line (ZPL) emission efficiency, which hinders the generation of indistinguishable photons. However, this is crucial for applications such as quantum memories and quantum computation.\newline
To overcome this challenge, group IV color centers have emerged as promising alternatives. Silicon\nobreakdash-vacancy centers \cite{Clark1995,Neu2011f,Sipahigil2014,Rogers2014,Schroder2017,Li2016} and germanium\nobreakdash-vacancy centers \cite{Nguyen2019,Nahra2021,Christinck2023} have shown potential as single\nobreakdash-photon sources due to their high ZPL emission efficiency \cite{Clark1995,Neu2011f,Sipahigil2014,Rogers2014,Schroder2017,Li2016}. These centers, however, suffer from short spin coherence times at cryogenic temperatures (around $4$\,K), requiring operation at dilution\nobreakdash-refrigerator temperatures (below several hundred millikelvin) to achieve long coherence times \cite{Sukachev2017, Senkalla2024}. More recently, lead\nobreakdash-vacancy centers have attracted interest due to their potentially long spin coherence times \cite{DTchernij2018, Trusheim2019,Wang2021,Wang2024}. However, the formation of these centers currently requires the application of high pressures and temperatures, such as $7.7$\,GPa and $2100^{\circ}$C.\newline
Tin\nobreakdash-vacancy (\ch{SnV}) centers are a promising solution to these issues. They exhibit a high Debye\nobreakdash-Waller factor of about $0.6$ \cite{Gorlitz2020} and maintain long spin lifetimes on the order of $10$\,ms at about $3$\,K without the requirement of a dilution refrigerator~\cite{Trusheim2020}. \ch{SnV} centers also show narrow linewidths of approximately $6$\,nm at room temperature \cite{Iwasaki2017,Bradac2019} and lifetime\nobreakdash-limited linewidths at low temperatures \cite{Trusheim2020,Rugar2020_NL,Gorlitz2020,Gorlitz2022}. Additionally, they possess capabilities as quantum electrometers for detecting localized charges at the lattice scale as was recently demonstrated \cite{Pieplow2024}. These characteristics make \ch{SnV} centers particularly attractive for photonic quantum technologies. Efficient photon extraction from \ch{SnV} centers in diamond is essential for practical applications. However, photon extraction is inherently limited due to total internal reflection caused by the high refractive index of diamond ($n \sim 2.4$). As a result, the saturation count rate detected from single \ch{SnV} centers in bulk diamond is typically limited to below $\sim 1$\,Mcps \cite{Tchernij2017}. To overcome this limitation, nanofabrication technologies have paved the way for creating nanostructures, such as one\nobreakdash-dimensional photonic crystal cavities with grating couplers \cite{Rugar2021}, diamond waveguides with tapered ends \cite{Martinez2022,Pasini2024}, and nanopillars \cite{Trusheim2020}. An alternative structure for the efficient extraction of photons are so\nobreakdash-called multi\nobreakdash-cones, tall nanopillars with tapered sidewalls and a multi\nobreakdash-cone parabolic structure \cite{Zhu2023}. Such curved structures enable the efficient collection of emitted photons, both upward and downward, using reflective surfaces or total internal reflection \cite{Schell2014b,Schell2014c}. Furthermore, the increased height and the multi\nobreakdash-cone structure contribute to enhanced extraction efficiency \cite{Zhu2023}. In addition to applications, such as quantum sources and quantum memories \cite{Ruf2021, Bhaskar2020, Knaut2024, Bersin2024}, this structure has attracted attention for sensing applications with high spatial resolution utilizing defect centers as probes for scanning probe microscopy \cite{Schell2014,Zhu2023,Pieplow2024}.\newline
Here, we report on the development of diamond multi\nobreakdash-cones containing single \ch{SnV} centers, demonstrating highly efficient single photon extraction. We demonstrate the successful fabrication of a pillar containing a single \ch{SnV} center that exhibits a sharp ZPL at approximately $619$\,nm with a full width at half maximum (FWHM) of $6$\,nm. Remarkably, the emitter exhbited an extraordinarily high saturation count rate of over $9$ million photons per second. Finite\nobreakdash-difference time\nobreakdash-domain (FDTD) simulations were performed to investigate the photon extraction from the nanopillars, revealing an extraction efficiency exceeding $70$\,\%, underlining the validity of our findings. These results establish our nanopillar platform as a promising candidate for bright and stable quantum sources and sensors based on \ch{SnV} centers in diamonds.

%
%
\section{Experimental section}
\subsection{Sample Fabrication}
We utilized a similar technique to fabricate the pillar structures containing \ch{SnV} centers as described in \cite{Zhu2023}. A diamond membrane with $20$\,µm thickness diced from an electronic\nobreakdash-grade diamond plate (element $6$) was used for this experiment. \ch{^{120}Sn} ions were implanted into this membrane with an implantation energy of $700$\,keV and an ion fluence of $5 \times 10^{10}$\,ions/cm\textsuperscript{2}. This implantation energy was selected to create a sacrificial layer to prevent the effect of surface oxidation during high\nobreakdash-temperature annealing in the next step. After the implantation, the sample was annealed at $1200$\,°C for $12$ hours in vacuum. 
The cones were lithographically defined using a layered electron\nobreakdash-beam resist, followed by electron\nobreakdash-beam lithography. Following the development, aluminum masks are deposited by electron\nobreakdash-beam evaporation and transferred onto the diamond by an inductively coupled plasma reactive\nobreakdash-ion etching (ICP RIE) process. After etching, the aluminum masks are removed in a triacid mixture.

\subsection{FDTD Simulations}
\begin{figure*}[ht!]
\centering
\hspace*{-0cm}
\includegraphics[clip, width=\textwidth,keepaspectratio]{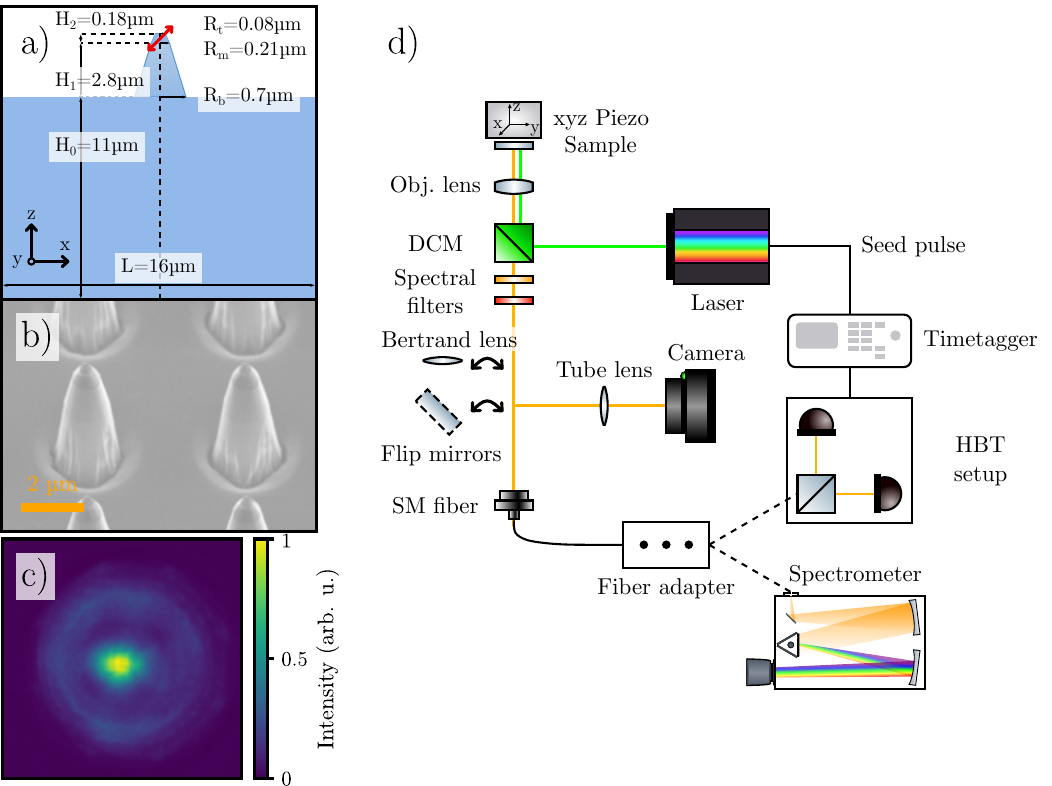}
\caption{Overview of the sample properties and experimental setup. a) Schematic illustration of the cone morphology used for the calculation model. b) SEM image of the multi-cone structures. c) Backfocal image of a single \ch{SnV} center inside a multi-cone. d) Schematic drawing of the fluorescence microscope setup.}
\label{fig:sample_setup}
\end{figure*}

In order to analyze the extraction efficiency, we performed three\nobreakdash-dimensional finite\nobreakdash-difference time\nobreakdash-domain (3D FDTD) optical simulations (Lumerical, FDTD solutions). The calculation model is shown in fig. \ref{fig:sample_setup} a). This model is designed based on the scanning electron microscope (SEM) images shown in fig. \ref{fig:sample_setup} a), as well as previous investigations of the pillar morphology \cite{Zhu2023}. A \ch{SnV} center was modeled as an electric dipole with an emission wavelength of $620$\,nm. The extraction efficiency to free\nobreakdash-space was calculated by normalizing the monitored power emitted by the dipole. Further details on the simulations and the results can be found in the supplementary material.

\subsection{Fluorescence Microscopy}
To probe the implanted defects we optically excite the sample and analyze the returning emission in a home\nobreakdash-built confocal microscope as shown in fig. \ref{fig:sample_setup} e). Pulsed laser light at $532$\,nm from a supercontinuum source (NKT, SuperK Fianium FIU15) is used for excitation. It is reflected by a long\nobreakdash-pass dichroic mirror and directed onto the sample by an air objective with a numerical aperture of $0.95$ (Olympus, MPLANPON100X). The sample is mounted on a nano positioner (PiezoSystemJena, Tritor 100) for position control. The fluorescence is collected in a multimode fiber and directed towards a Hanbury\nobreakdash-Brown\nobreakdash-Twiss interferometer consisting of two avalanche photodiodes (Laser Components, Count-100C) or a spectrograph (Princeton Instruments, SpectraPro HRS500), respectively. We perform each measurement under the usage of different spectral filters in order to investigate the background contributions from different spectral ranges.\newline
To investigate the overall density of centers inside the cones as well as the distribution of different emission lines of \ch{Sn} related defects, we measure a total of $126$ cones from an $80$\,$\times$\,$80$\,µm\textsuperscript{2} scan range. To do so in a reproducible way, the cones are selected and successively measured in a fully automatic manner. Details about this process are found in section $4$ of the supplementary information. For each cone, the emission spectrum, count rate, autocorrelation and lifetime are measured under $532$\,nm excitation at $120$\,µW. All data is analyzed automatically as well in order to guarantee an objective evaluation of the implantation process.
\begin{figure*}[ht!]
\centering
\hspace*{-0cm}
\includegraphics[clip, width=\textwidth, keepaspectratio]{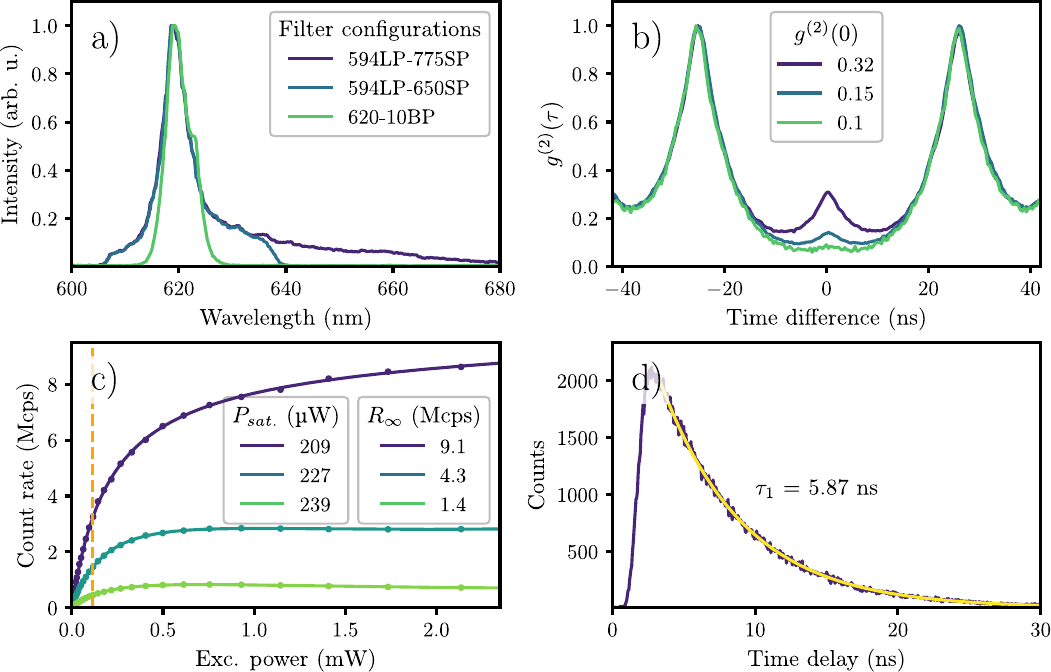}
\caption{Properties of a single \ch{SnV} center in a multi cone measured across different spectral ranges. a) shows the normalized emission spectra measured with three different sets of spectral filters. Pronounced emission at $619$\,nm indicates the presence of the \ch{SnV^{-}} center. b) and c) show the autocorrelation and saturation measurement for each filter configuration, respectively. The autocorrelation measurement is conducted under $532$\,nm exciation at $120$\,µW. This particular power is marked in c) by the dashed orange line. A lifetime measurement is shown in d) with a lifetime of $5.87$\,ns retrieved from an exponential fit (yellow line). Only a single measurement is depicted as there were no differences in the data for the different used spectral filters.}\label{fig:single_cone}
\end{figure*}

%
%
\section{Results and discussion}
To verify the general feasability of the cone structures and the implantation procedure, we first present results of a single fluorescent center inside one of the cones. Figure \ref{fig:single_cone} shows an overview of the performed measurements. All measurements are repeated under usage of different combinations of spectral filters. Purple, teal and green data indicate the used configurations $594$\,nm longpass and $775$\,nm shortpass, $594$\,nm longpass and $650$\,nm shortpass, and a $620$\,nm bandpass filter with a full width half maximum of $10$\,nm, respectively. The combinations of these different filters will be referred to in the following as configuration $1$, $2$, and $3$, respectively. It is worth noting that the bandpass filter has a considerably lower transmission maximum ($60$\,\%) compared to the other filters, which explains the notably reduced emission count rates in this setting. Panel a) shows the resulting normalized fluorescence emission spectrum for the different implemented filters. The pronounced zero phonon line at $619$\,nm indicates that this particular cone contains the negatively charged \ch{SnV^{-}} center \cite{Bradac2019,Gorlitz2020}. Panel b) shows the second-order autocorrelation measurement recorded at $120$\,µW and $39$\,MHz excitation. The saturation measurements for the different filter settings are given in c), with the data points indicated by dots. The solid lines represent fit functions to the recorded data and the dashed vertical line highlights the power at which the autocorrelation measurements in b) are acquired. In particular, the two measurements in configurations $2$ and $3$ (teal and green curve) exhibit a decrease in the collected light for high excitation power, which could be signs of a non\nobreakdash-trivial energy level dynamic. Thus the data is fitted as a function of excitation power $P$ by a modified formula
\begin{equation}
R(P) = R_{\infty} \frac{P}{P + P_{sat.} + c_{sh.} \cdot \frac{P^2}{P_{sh.}}} + c_{bg.} \cdot P,
\end{equation}
similar to \cite{Nahra2021}, with the saturation power $P_{sat.}$ and the fluorescence emission at infinity $R_{\infty}$. The parameters $c_{bg.}$, $c_{sh.}$ and $P_{sh.}$ phenomenologically describe the influence of background fluorescence and the observed shelving of the curve, respectively. Lastly, a lifetime measurement is shown in d) measured with the bandpass filter around the zero phonon line. The solid yellow line shows an exponential fit function to the data (purple). For the lifetime measurements, virtually no difference could be observed under usage of different filters. Thus, the other measurements are omitted. This particular cone shows irrefutable quantum emission with a $g^{(2)}(0)$ value of $0.32$ across the full measured emission range, which signals a predominant single\nobreakdash-photon character of the fluorescence. Simultaneously, a count rate of over $3$\,Mcps is achieved as read off of the saturation measurement. The fit to the saturation reveals a relatively low saturation power of $209$\,µW and count rates of $9.1$ million counts per second at infinity. However, a non\nobreakdash-negligible linear contribution in the saturation from the background fluorescence of the diamond is present, which is also reflected in the deviation of the $g^{(2)}(0)$ value from $0$. When restricting the detected emission by use of the $650$\,nm shortpass filter or the bandpass filter around $620$\,nm the $g^{(2)}(0)$ value improves to $0.15$ and $0.1$, respectively. While substantially dimmer, the emitter still shows saturation count rates of $4.3$ and $1.4$ million counts per second, respectively. The fit function to the lifetime measurement reveals a time constant $\tau = 5.87$\,ns associated with the lifetime of the excited state. This result is in good agreement with existing literature on \ch{SnV} centers in bulk diamond \cite{Iwasaki2017, corte2022}.\newline
Our results are compared to other studies in table \ref{tab:snv_overview}. It can be seen that while the single\nobreakdash-photon purity leaves room for improvement, the brightness of the source is greatly enhanced when embedded into the multicone structure. In the limited spectral range our results still show a significant increase in brightness while lowering the $g^{(2)}(0)$ value to a comparable level.\newline
Additionally, a backfocal image of the emitter is shown in figure \ref{fig:sample_setup} c). The densely concentrated area of high intensity at the center indicates a substantial contribution to the propagated light by a nearly Gaussian mode. This particular part of the emission can most likely be attributed to the mode profile of the cone. Consequently, this observation can be considered an optimistic evaluation of the ion implantation process regarding both implantation depth and position.
\begin{table*}[h!]
  \begin{center}
\begin{tabular}{|c|c|c|c|}
\hline
Publication & Sample & Count rate $R_\infty$ & $g^{(2)}(0)$\\
\hline
\hline
This work & diamond membrane & $\approx 9$\,Mcps & $\approx 0.3$\\
 & \& multi-cones & &\\
\hline
Takashima et. al. \cite{Takashima2025}& nanodiamond & $492$\,kcps & $0.04 \pm 0.11$ \\
\hline
Tchernij et. al. \cite{Tchernij2017} & bulk & $1.37 \pm 0.11$\,Mcps & $0.29 \pm 0.02$ \\
\hline
Iwasaki et. al. \cite{Iwasaki2017} & bulk & $530$\,kcps & $\approx 0.5$\\
\hline
Görlitz et al. \cite{Gorlitz2020} & bulk & $80-150$\,kcps & $0.05$ \\
\hline
Corte et al. \cite{corte2022} & bulk & $185$\,kcps & $<0.5$ \\
\hline
\end{tabular}
  \end{center}
\caption{Comparison of the achieved benchmark properties from single \ch{SnV} centers across the existing literature and in this work. The measured saturation count rate of $9$\,Mcps greatly surpasses all other reported sources, while a clear single-photon signature is maintained.}\label{tab:snv_overview}
\end{table*}
\begin{figure}[ht!]
\includegraphics[clip, width=\textwidth, keepaspectratio]{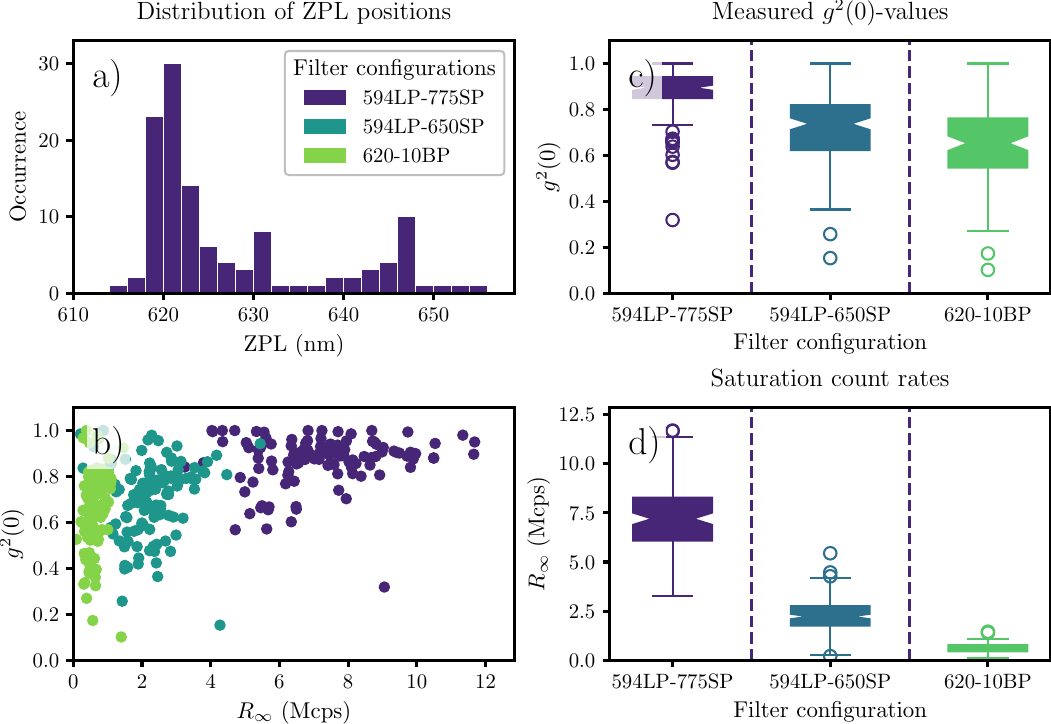}
\caption{Overview of the measured properties across all cones for different spectal filter configurations. a) Distribution of ZPL positions. The majority of observed ZPLs are located around $620$\,nm. 
b) Scatter plot of the measured $g^{(2)}(0)$ values over the count rates in saturation. The different filter settings are coded by different colors. c) and d) Boxplots of the  $g^{(2)}(0)$ values and saturation count rates, respectively. The $g^{(2)}(0)$ values are improved, while the count rates in saturation are drastically reduced by the restriction of the measured spectral range.}
\label{fig:statistics}
\end{figure}
\newline
The evaluation of the $126$ measured cones reveals that the majority of cones do host fluorescent defects with observable antibunching. The relevant measured quantities for all investigated cones are recapitulated in figure \ref{fig:statistics}. A histogram of the exctracted zero phonon lines from the measured emission spectra with a binning of $2$\,nm is shown in a).\footnote{If multiple significant peaks are present in a recorded emission spectrum the most dominant one is registered as the ZPL.}. 
b) shows a scatter plot of the measured $g^{(2)}(0)$\nobreakdash-values over the saturation count rates obtained from the fit functions. The panels c) and d) depict these two properties as box plots, respectively. Again, all measurements are repeated for all three filter configurations as in the previous example. The color coding indicates the full spectral range by purple, the moderately restricted range by teal, and the narrow range around $620$\,nm by green data. The latter spectral ranges are omitted for the ZPL histogram in order to maintain a clear overview. The distribution of ZPLs shows a dominant peak between $615$\,nm and $625$\,nm (see fig. \ref{fig:statistics}) for $80$ out of the $126$ cones. Further peaks in the histogram can be observed around $630$\,nm and $645$\,nm. These lines have been reported before with a direct relation to \ch{Sn} based defects, but the exact related atomic configuration is as of yet unclear \cite{Tchernij2017,Iwasaki2017,Alkahtani2018,Luhmann2019,Rugar2020_NL,Luhmann2020,corte2022,Luhmann2020}. All other contributions are insignificant.\newline
For the measurements in filter configuration $1$ we find $g^{(2)}(0)$ values between $0.3$ and $1$, with half of them lying between $0.85$ and $0.95$. These values indicate that while the majority of cones show obvious signs of quantum emission, it cannot be attributed to single centers in most cases. The corresponding count rates in saturation vary from $3$ to $11.5$ million counts per second. Restricting the emission range comes with an expected improvement of the measured $g^{(2)}(0)$ values and a stark decrease of the saturation count rates. For the second and third filter configuration the $g^{(2)}(0)$ values range from $0.15$ to $1$ and from $0.1$ to $1$, respectively. The median (mean) values improve from $0.87$ ($0.89$) over $0.74$ ($0.71$) to $0.65$ ($0.64$) throughout the different settings. The count rates in saturation span from $0.2$\,Mcps to $5.4$\,Mcps and from $0.09$\,Mcps to $1.5$\,Mcps for filter settings two and three. These results are cohesive, as the the spectral distribution among measured cones suggests that the constraint in the measured emission interval removes the identified centers at longer wavelengths. Hence the drop in brightness and increase in single photon purity is expected. Simultaneously, the fact that most measured $g^{(2)}(0)$ values remain above a threshold of $0.5$ in all settings proofs that most cones hold at least two or more \ch{SnV^{-}} centers associated with the emission around $619$\,nm.

%
%
\section{Conclusion}
Our experimental results demonstrate the superb capability of the diamond multicone structures to enhance the collection efficiency of \ch{SnV} defect centers. We achieved narrow emission (FWHM of $6$\,nm) centered around $619$\,nm with a high saturation count rate of approximately $9$\,Mcps, far exceeding the results of other studies.  Furthermore, the second-order correlation function exhibited an antibunching dip well below $g^{(2)}(0) = 0.5$, indicating single-photon emission. Overall, our study convincingly shows the successful incorporation of fluorescent \ch{SnV} centers into diamond multicone structures for enhanced collection efficiency. Our findings further suggest that the majority of cones host fluorescent defects with clearly visible non\nobreakdash-classical emission, but with a $g^{(2)}(0)$ value above $0.5$. This indicates that the irradiation dose still favors the creation of multiple emitters per cone. With small adjustments to the implantation process, our method could pave the way towards the reliable utilization of these promising quantum emitters with greatly improved photon collection capabilities.

\section{Acknowledgements}
We gratefully acknowledge financial support from JST PRESTO (grant no. JPMJPR2257), JST ERATO TAKEUCHI Super Quantum Entanglement program (grant no. JPMJER2402), JSPS KAKENHI (grant nos. 24H00195, 21H04444, 26220712, and 23K22426), MEXT Q-LEAP (grant no. JPMXS0118067634), the JSPS Bilateral Program (grant no. JPJSBP120242003), the Murata Science and Education Foundation, the Matsuo Foundation, SUHARA Memorial Foundation, and a Research Fellowship from the Japan Society for the Promotion of Science (grant no. 23KJ1190). 
A part of this work was supported by ARIM Project of the Ministry of Education, Culture, Sports, Science and Technology (MEXT)  through grant no. JPMXP1224AT0335. This work was also funded by the Deutsche Forschungsgemeinschaft (DFG, German Research Foundation) under Germany’s Excellence Strategy within the Cluster of Excellence QuantumFrontiers (EXC 2123, Project No. 390837967) and within the Cluster of Excellence PhoenixD (EXC 2122, Project No. 390833453). We thank Konstantin Herb for suggesting a diamond nanopillar structure for \ch{SnV} centers.

\section{Author contributions}
P. Tieben performed the measurements, evaluated the results, and wrote the manuscript with contributions from all co-authors. H. Takashima performed the initial experiments and the FDTD simulations, and contributed to the design of the experiment and the evaluation of the results. J. Rhensius designed, fabricated and analyzed the sample. T. Segawa provided input on the measurement results and data analysis, and contributed to the review and revision of the manuscript. R. Abe performed the initial FDTD simulations. H. Shimazaki performed the initial experiments. S. Takeuchi provided the experimental apparatus for initial experiments. A. W. Schell designed and supervised the study. All authors discussed and interpreted the data.

\clearpage
\bibliography{bib-SnV}
\end{document}



\section{FDTD Simulation of Extraction Efficiency and Far-Field Intensity}
\begin{figure}[t]
\begin{center}
\includegraphics[clip,width=0.75\textwidth,keepaspectratio]{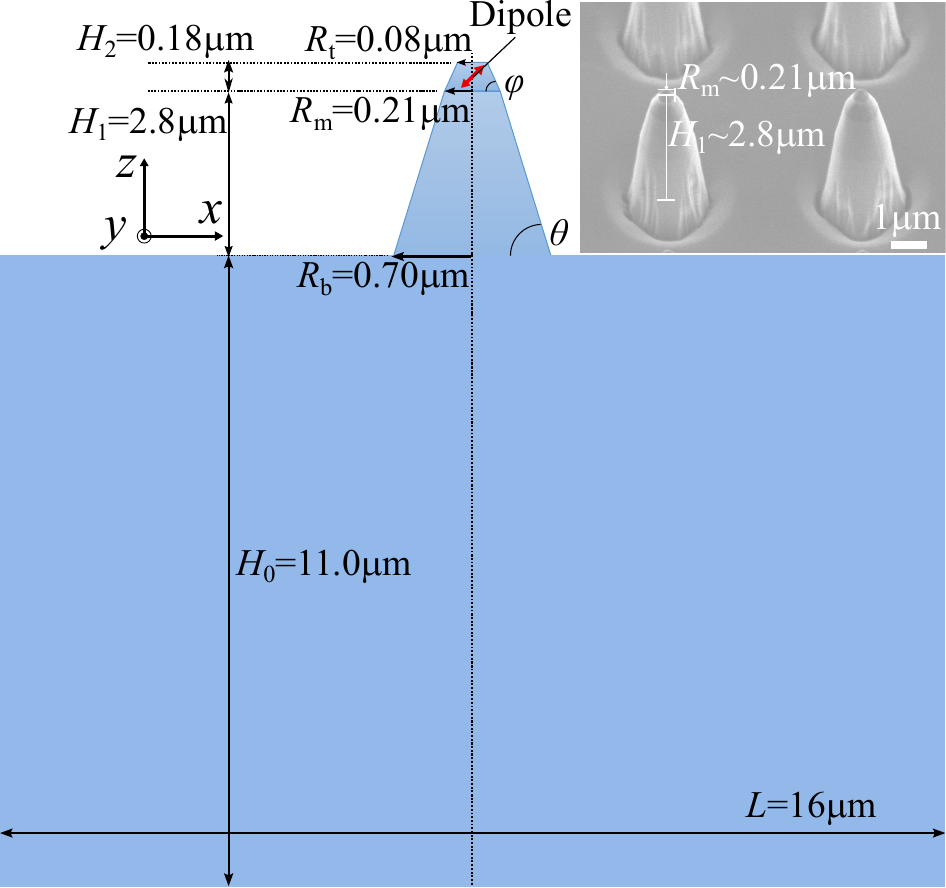}
\caption{Calculation model for FDTD simulation. The inset is an SEM image used to estimate the parameters.}
\label{fig:S1}
\end{center}
\end{figure}

\begin{figure}[t]
\begin{center}
\includegraphics[clip,width=0.85\textwidth,keepaspectratio]{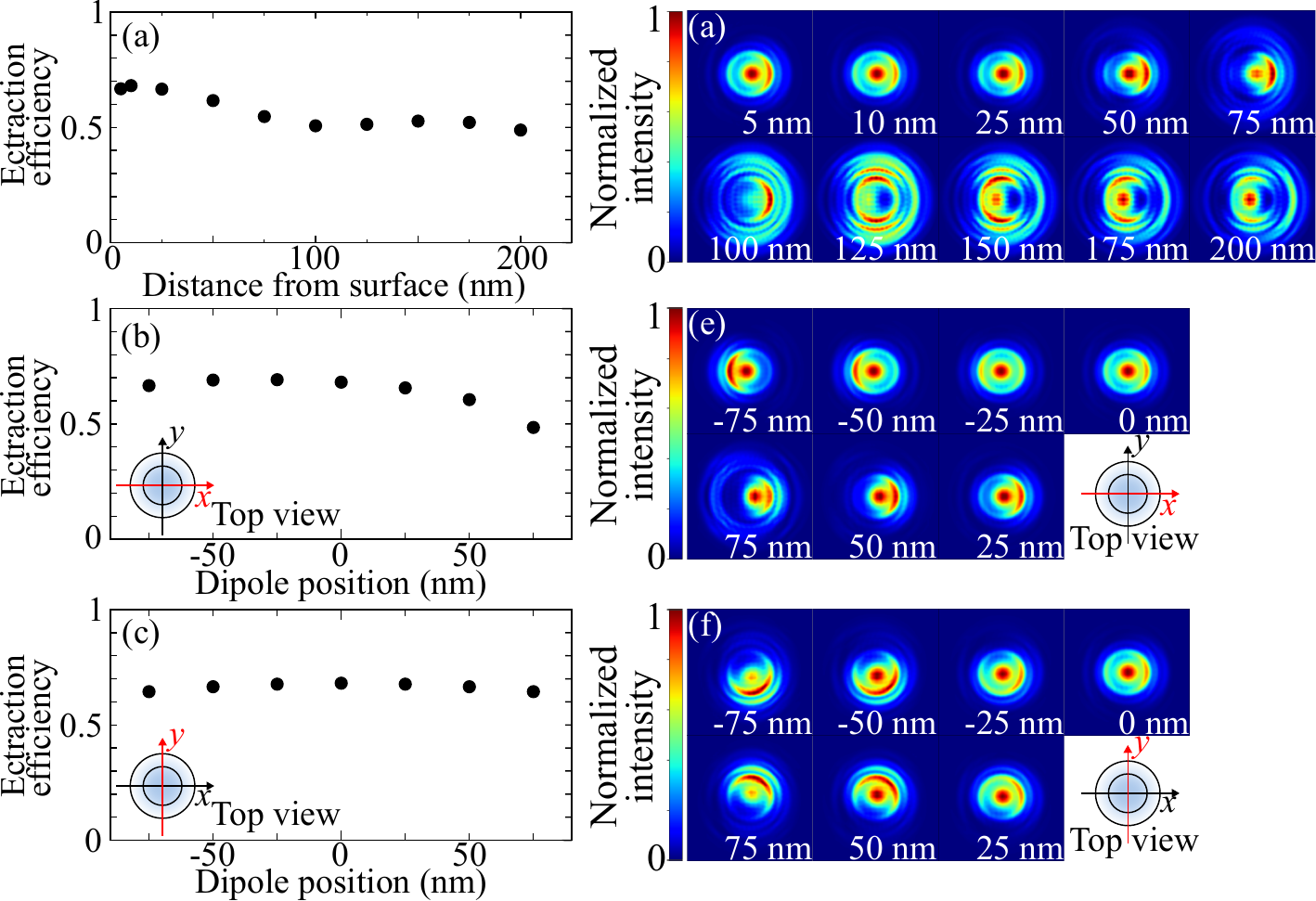}
\caption{(a), (b), and (c) Calculated extraction efficiencies when the dipole is moved along the $z$-axis (a), $x$-axis (b), and $y$-axis (c), respectively. (d), (e), and (f) Calculated far-field patterns when the dipole is moved along the $z$-axis (d), $x$-axis (e), and $y$-axis (f), respectively.}
\label{fig:S2}
\end{center}
\end{figure}

Here, we describe the results of our FDTD simulations for the extraction efficiencies, as well as the far-field patterns. A schematic illustration of our calculation model is depicted in figure \ref{fig:S1}. The exact parameters of this model are partly observed directly from SEM analysis of the fabricated pillars, as well as estimations based on the fabrication procedure and previous investigations of the cone geometry \cite{Zhu2023}. Specifically, the height of the lower region of the pillar ($H_1$), height of the upper region of the pillar ($H_2$), the bottom radius ($R_{\mathrm{b}}$), and the middle radius ($R_{\mathrm{m}}$) were estimated directly from the SEM analysis of the fabricated pillars. The sidewall angles ($\theta = 80^{\mathrm{o}}$, $\phi = 54.5^{\mathrm{o}}$) were estimated from the angles reported in previous investigations\cite{Zhu2023}. The substrate thickness ($H_0$) was based on the surface treatment procedure \cite{Zhu2023}. The final parameters were $H_1 = 2.8$\,µm, $H_2 = 0.18$\,µm, $R_{\mathrm{b}} = 0.7$\,µm, $R_{\mathrm{m}} = 0.21$\,µm, $R_{\mathrm{t}} = 0.08$\,µm, and $H_0 = 11.0$\,µm.  The \ch{SnV} center was modeled as an electric dipole aligned along the ($111$) direction of a ($100$) plane with an emission wavelength of $620$\,nm. The calculation area for the FDTD simulations was set to $16 \times 16 \times 16.5$\,µm\textsuperscript{3}. The optical power emitted by the dipole is monitored at the bottom edge of the substrate. The extraction efficiency can then be calculated by normalizing the monitored power with respect to the total emitted power. The far-field patterns are also calculated as the far-field projection of the emission intensity at the bottom edge of the substrate.
\newline
First, we investigate the behavior of the extraction efficiencies for vertical movement of the dipole along the $z$\nobreakdash-axis in the range from $5$ to $200$\,nm below the tip of the pillar, as shown in Figure \ref{fig:S2}(a). The extraction efficiency reaches a maximum of $68$\,\% when the dipole is positioned $10$\,nm below the surface. It then decreases to $50$\,\% efficiency when the dipole is $100$,nm below the surface, slightly increases to $53$\,\% at $150$\,nm, and decreases again to $49$\,\% at $200$\,nm below the surface. Next, the dipole is moved in the $x$\nobreakdash-$y$\nobreakdash-plane. For this scenario, the dipole is fixed at a depth of $10$\,nm to maximize the efficiency. Figure \ref{fig:S2}(b) and (c) show the simulated extraction efficiencies when the dipole is moved laterally. The dipole is moved along the $x$\nobreakdash-axis from $-75$\,nm to $+75$\,nm around the symmetry axis of the cone. The calculated extraction efficiency is highest around $69$\,\% when the dipole is located between $-25$ and $-50$\,nm from the center. It is asymmetrically reduced and experiences a steep reduction along the positive $x$ direction, likely due to the dipole tilting in the positive direction of the axis, causing increased light leakage as it approaches the end of the pillar. Along with the $y$ direction, the extraction efficiency behaves symmetrically due to the dipole orientation. Overall, it varies only marginally with a reduction to a minimum of $64$\,\% at $\pm 75$\,nm.\newline
The simulated far\nobreakdash-field patterns reveal the critical dependence on position within the cone for the propagated mode profile. The results are shown in figure \ref{fig:S2} (d), (e), and (f). All plots are normalized with respect to the maximum value of each graph. Along the $z$\nobreakdash-axis, the emitted intensity is highly confined and relatively symmetric only for depths around $10$\,nm. For larger depths, secondary patterns at greater distances from the optical axis form, and the overall intensity is spread out.\newline
For movement in the $x$\nobreakdash-direction, the pattern is most confined for a slightly off\nobreakdash-centered dipole around $-25$\,nm. The intensity distribution forms additional asymmetric patterns when it is moved away from this position. Along the $y$\nobreakdash-axis, the emission pattern is relatively symmetric at the center position. Away from this position, a side lobe in the intensity pattern can be observed that rotates clock\nobreakdash-wise and counter clock\nobreakdash-wise for movement in the negative and positive direction, respectively.\newline
Overall, our simulations suggest that the emission of an \ch{SnV} center can be collected most efficiently if the center is implanted close to the tip of the cone structure, around $10$\,nm below the surface, and almost perfectly centered along its symmetry axis. Depending on the tilt of the dipole, a slight offset from the center of about $25$\,nm is favorable for extraction from the diamond substrate. 

\section{Fluorescence Emission Line Width}
\begin{figure*}[ht]
\hspace*{8,8mm}
\includegraphics{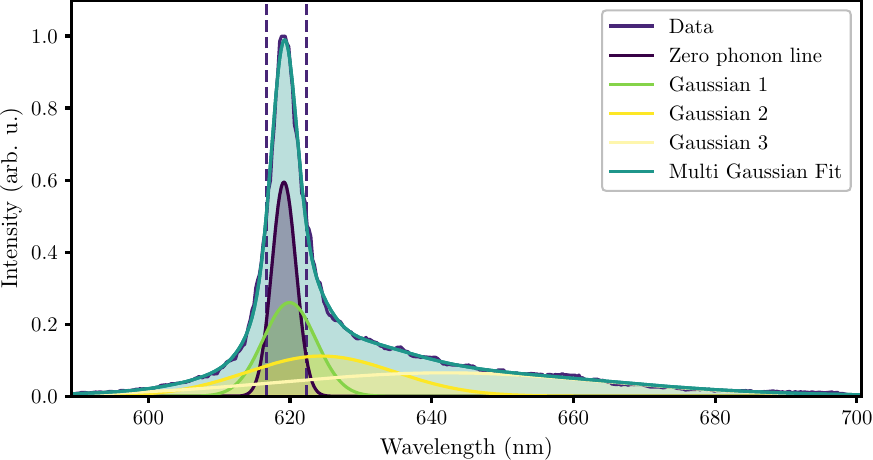}
\caption{Multi-Gaussian fit of the fluorescence emission spectrum of a single $SnV^{-}$ center. The shaded outlined areas show the individual Gaussian contributions and the full fit function (teal) to the data. The spectrum is recreated by a sum of four separate Gaussian functions. The resulting zero phonon line (purple) exhibits a width of $4$\,nm. The full width half maximum of $5.6$\,nm is depicted by the dashed vertical lines for comparison.}
\label{fig:spectrum_fit}
\end{figure*}
To analyze the spectral features of the emitters, we fit the emission spectrum by a sum of multiple Gaussian distributions. In this way, the contributions of the zero-phonon line as well as several different phonon modes can be retrieved. Such a fit for the emission spectrum of the presented emitter from the main text is shown in figure \ref{fig:spectrum_fit}. The best fit result is obtained for a combination of four Gaussian contributions. The ZPL at $619.14$\,nm (purple) is broadened significantly by a low\nobreakdash-energy component (green) with a detuning of only $2.5$\,meV. The phonon sideband is made up of two components at $17$ (yellow) and $71$\,meV (light yellow) detuning from the ZPL, respectively. The fitted line width of the ZPL is extracted as $4.0$\,nm in this way. In contrast, the full width at half maximum (FWHM) (compare dashed vertical lines) of the main peak yields a value of $5.6$\,nm.

\section{Multiple Wavelength Excitation}
\begin{figure*}[ht]
\hspace*{8,8mm}
\includegraphics{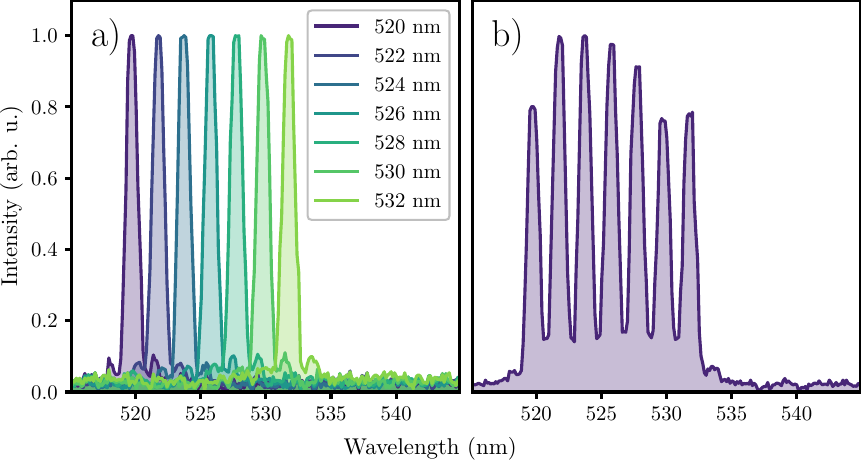}
\caption{Measured output spectra of the supercontinuum laser. a) shows seven different spectra for single selected wavelengths from $520$ to $532$\,nm with a step size of $2$\,nm. All spectra are normalized individually with respect to the maximum intensity. b) shows the combined output spectrum when all wavelengths from a) are selected simultaneously.}
\label{fig:multiple_wavelengths}
\end{figure*}

Due to the limited spectral power density of the supercontinuum laser, i.e., the limited output power for single selected wavelengths, we use the capability of the acousto\nobreakdash-optic filter to combine multiple wavelengths simultaneously. In accordance with the linewidth of about $1$\,nm for a single selected wavelength, the total power for multi\nobreakdash-wavelength operation can be maximized by a spacing of $\sim 2$\,nm or more. For the saturation measurements, we use seven different excitation wavelengths simultaneously from $520$ to $532$\,nm in $2$\,nm steps. Figure \ref{fig:multiple_wavelengths} a) shows the laser spectrum for the single wavelengths, normalized to maximum intensity. The resulting output spectrum for the combined wavelength operation is shown in b). Note that the overall spectral density of the supercontinuum source greatly varies with wavelength, resulting in the different intensities of the individual contributions in b). The ratio of the different wavelength contributions is fixed with respect to the achieved ratio, when all involved wavelengths are set to their maximum output. This way the spectral distribution can be kept constant across the full range of possible output powers. By this approach, we are able to generate a total output power of $3.5$\,mW ($2.1$\,mW) at a repetition rate of $78$\,MHz ($39$\,MHz). This corresponds to an increase in power by a factor of $8.75$ over the maximum power of $400$\,$\mu$W ($240$\,$\mu$W) for single wavelength operation at $532$\,nm. The stated excitation powers in the presented saturation measurements refer to the total output power of all combined wavelengths in this context.

\section{Automatic Cone Selection}
\begin{figure*}[ht!]
\hspace*{21,3mm}
\includegraphics{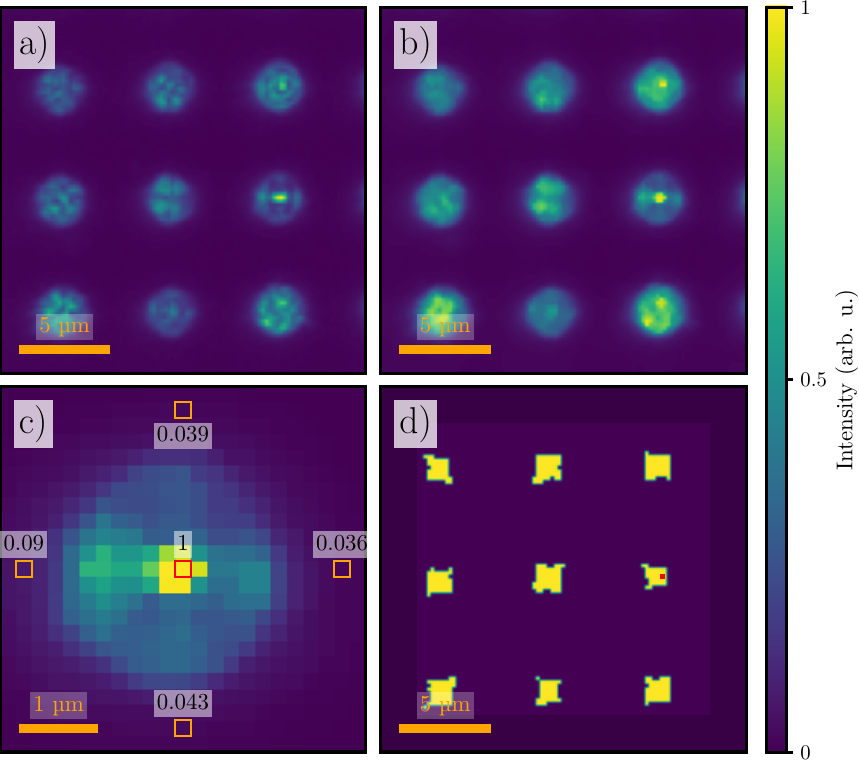}
\caption{Overview of the automatic cone selection process. a) Representative $20 \times 20$\,µm\textsuperscript{2} fluorescence scan of the sample. b) The same scan with a Gaussian filter applied to it. c) Illustration of the selection criterion for an exemplary pixel. The brightness of the pixel in the red square is compared to all the pixels in the orange squares and is selected, if it exceeds each of them by a chosen factor. d) Selected pixels (yellow) based on c) for the full scan. Connected pixels are grouped together and interpreted as one cone. The exemplary pixel from c) is highlighted and the shaded edges indicate the areas that are exempt from the selection process as explained in the text.}
\label{fig:pixel_selection}
\end{figure*}
The emitter selection process for the series of automated measurements is performed in the same way as in \cite{tieben2024}. Single pixels of increased brightness are selected from $20 \times 20$\,µm\textsuperscript{2} fluorescent scans (fig. \ref{fig:pixel_selection} a) as follows. The scan is first subjected to a Gaussian filter with a kernel size of three by three pixels (fig. \ref{fig:pixel_selection} b). From the resulting image, each pixel is evaluated by comparing its brightness to the $\pm n$\textsuperscript{th} neighboring pixel in $x$- and $y$-direction. If its brightness exceeds the brightness of all $n$\textsuperscript{th} neighbors by a variable value of $a$, it is selected. For this sample, the pixel size has been chosen as $0.2 \times 0.2$\,µm\textsuperscript{2} and the values for $n$ and $a$ have been chosen as $10$ and $2.5$, respectively. An example of the selection process is shown in figure \ref{fig:pixel_selection} c). The red outlined pixel at the center is compared to its $10$\textsuperscript{th} neighbors in each direction (orange outline). Their relative brightness with respect to the center pixel is included alongside. Since the tested pixel exhibits a brightness of more than the chosen $2.5$ times the value of all other pixels, the condition is fulfilled. The last panel (d) shows the result of the procedure for the full $20 \times 20$\,µm\textsuperscript{2} area. Yellow pixels indicate the selected pixels while the rest is left purple. The dark shaded edges mark the regions that are not considered in this method because the $10$\textsuperscript{th} neighbor does not exist for all directions in these regions. This fact automatically avoids twin measurements for cones sitting on the edges of two adjacent scans. After the selection of bright pixels from the scan, adjacent pixels (along $x$-, $y$-axis, and diagonally) are grouped together and interpreted as a single bright island. For the performed measurement sequence, the brightest point of each identified cone is chosen as an initial position, followed by an \textit{xyz}\nobreakdash-optimization step.

\clearpage
\bibliography{bib-SnV}